# High Q-factor Fano resonances in coupled wire arrays with bulk structural asymmetry

David E. Fernandes[1*], Sylvain Lannebère[2], Tiago A. Morgado[2]

[1]Instituto de Telecomunicações, Avenida Rovisco Pais, 1, 1049-001 Lisboa, Portugal

[2]Instituto de Telecomunicações and Department of Electrical Engineering, University of Coimbra, 3030-290 Coimbra, Portugal

## Abstract

In this article we characterize the macroscopic electromagnetic response of a nested wire metamaterial with a high degree of bulk structural asymmetry. The unit cell of the considered metamaterial contains two sets of metallic wires, one set consisting of an array of straight wires and another formed by a racemic array of helical-shaped wires. We study the scattering of electromagnetic waves in a metamaterial slab and show that the electromagnetic coupling between both arrays of wires can originate Fano resonances with narrow lineshape. The origin of the resonances is rooted in the formation of a sub-radiant mode in the metamaterial wherein the net polarization vector vanishes. We envision that the proposed configuration, with sharp resonances whose quality factor can be greatly enhanced by tuning the structural parameters of the wires, may have promising applications in sensing and switching in a broad frequency range from the microwave regime up to THz frequencies.

[*] To whom correspondence should be addressed: E-mail: dfernandes@co.it.pt

# I. Introduction

The uniaxial wire medium consists of a set of thin infinitely long wires oriented along the same direction and embedded in a dielectric medium [1-2]. This class of metamaterials has been extensively studied over the last two decades due to their unique electromagnetic properties such as a hyperbolic dispersion [3], a non-local (spatially dispersive) response [4-6], an anomalously high density of photonic states [7-8], amongst others [9]. Such remarkable properties pushed the wire medium into the spotlight of metamaterials research, allowing for numerous applications and exotic wave phenomena from microwave regime to infrared and optical frequencies [10-19].

Throughout the years, several proposals have emerged aiming to extend the range of applications for the wire medium configuration through various structural modifications. Among these modifications perhaps the most well-known correspond to inserting metallic plates in the wires [20-24], having multiple perpendicular arrays of connected and non-connected wires [25-29], or even by twisting the wires so that they become helical-shaped [30-37]. Quite interestingly, a wire metamaterial formed by a racemic array of helical-shaped wires can provide for unique opportunities [36-38]. In particular, it was shown in [36] that a racemic array of helical-shaped wires behaves as a hyperbolic material with an indefinite electric local response that enables negative refraction and partial focusing for transverse magnetic (TM) polarized waves. Furthermore, it was theoretically [37] and experimentally [38] demonstrated that a racemic helical-shaped wire metamaterial can behave as a magnetic analogue of the standard uniaxial wire medium, enabling the channeling of the subwavelength details of transverse electric (TE) polarized waves.

On the other hand, in [39] we showed that by having two straight wire arrays oriented parallel to each other in the same unit cell can allow to further manipulate the scattering properties of the metamaterial. We named this metamaterial nested wire medium and demonstrated that the scattering properties of the metamaterial can have strong Fano-type resonances. Fano resonances were discovered over fifty years ago by Ugo Fano [40] when studying the autoionizing states of atoms. Resulting from the interference of a discrete (dark) state with a broad (bright) band of continuum states [40, 41], Fano resonances have a characteristic narrow asymmetric lineshape. Fano resonances can have important applications [42] namely in the development of novel sensors and filters [43-45] and can also be used in the context of optomechanical interactions [46, 47]. We demonstrated in Ref. [39] that the interference of states in the nested wire metamaterial may be achieved by introducing some sort of structural asymmetry in the configuration. Fano resonances have also been previously studied in several other plasmonic structures and metamaterials [41,48-52].

In this work we propose a new nested wire metamaterial configuration with structural asymmetry to obtain Fano resonances. In this metamaterial configuration both sets of wires are made of the same material and are simply severed at the interfaces. The asymmetry required to obtain Fano-type resonances is introduced in the bulk metamaterial. Specifically, we consider one sub-array of straight wires while the other is formed by a racemic array of helical-shaped wires. Based on the homogenization formalism put forward in [39], we propose an effective medium model to characterize the response of the bulk nested wire metamaterial. Using this model and numerical simulations, we demonstrate that a slab of the nested helical wire medium can exhibit sharp Fano resonances with extremely high quality ($Q$) factor that are largely insensitive to the incidence angle of the excitation waves.

At this point it is important to highlight the main differences between the metamaterial proposed in this work and the nested wire configuration studied in [39]. As previously discussed, the metamaterial studied in [39] allowed to obtain Fano resonances based on structural asymmetries, either by having a slab terminated differently for each set of wires (geometrical-type asymmetry), or having each sub-array of wires made of different conducting materials (material-type asymmetry). In the first case, the asymmetry was associated with using metallic plates terminating only one set of wires. Hence, such geometrical-type asymmetry was linked to a broken structural symmetry at the interfaces rather than in the bulk, as in the configuration proposed in this work. Furthermore, the bulk structural asymmetry here is independent of the conducting material used in the wire arrays, contrary to the case of the material-type asymmetry discussed in [39]. For a material-type asymmetry the responses of the metallic wires of each array must be sufficiently different for the Fano resonances to emerge. However, finding conductive materials with responses distinct enough may be quite challenging, particularly in the microwave and millimeter wave frequency range where the metals behave as perfect electric conductors (PEC). The differences between the nested wire metamaterial studied in this work and the configurations in [39] may also be explained using their equivalent circuits. A nested wire metamaterial featuring two sub-arrays of wires can be seen as a combination of two LC series circuits, with each sub-array corresponding to one LC series circuit (please refer to Ref. [23] for further details), interconnected in a parallel configuration. Fano resonances emerging from the interaction between LC circuits have previously been discussed in [53]. In the configurations explored in reference [39], the Fano resonances were induced by altering the conducting material within one of the sub-arrays of wires, which consequently modified the per unit length inductance of the wires. Alternatively, these resonances were also generated from an interface-based geometrical-type asymmetry, in which a metallic patch introduced an additional capacitance in the corresponding LC series circuit. In this work, we combine an array of straight metallic wires with another array of helical-shaped metallic wires to create a bulk structural-type asymmetry. The helical-shaped wires enable simultaneous manipulation of wire inductance and capacitance, and thus may offer a more comprehensive approach to achieve and control Fano resonances. In summary, in this manuscript we propose a solution that emphasizes a bulk geometrical-type asymmetry which is not reliant on specific conducting materials or specific terminations of the slab, making it potentially more practical and applicable in certain frequency ranges

This paper is organized as follows. In Sec. II we propose a homogenization model to study the effective response of the bulk nested helical wire medium formed by one array of straight wires and one racemic array of helical-shaped wires. In Sec. III we present numerical simulations of the scattering properties of a metamaterial slab that validate the proposed effective medium model. The theoretical and numerical calculations highlight the unique scattering properties of the metamaterial, revealing the emergence of Fano resonances resulting from the interaction between both sets of wires. We also highlight that the quality factor of these resonances can be greatly enhanced by the proper tuning of the structural parameters. Finally in Sec. IV the conclusions are drawn.

## II. Effective medium response of the nested wire metamaterial

In this work we study the electromagnetic response of the nested wire metamaterial formed by two different sets of wires wherein in each square unit cell there is one set of straight wires and another set formed by four helical-shaped wires (two right-handed and two left-handed helical-shaped wires) arranged in a checkerboard pattern [see Fig. 1]. In what follows we rely on effective medium theory to characterize the electromagnetic response of the wire metamaterial structure [29, 39, 54-58]. In particular, based on the ideas discussed in previous works [29, 39], the effective permittivity of a nested wire metamaterial formed by $l=1,...,N$ sub-arrays of wires can be determined from the response of each individual sub-array provided all arrays are not strongly coupled in the near-field. In that case the influence of each sub-array on the other can be treated as a macroscopic excitation and its contribution to the macroscopic electric polarization vector **P** can be expressed as:

$$\mathbf{P}_l = \left[ \bar{\bar{\varepsilon}}_{\text{eff},l} - \varepsilon_h \varepsilon_0 \bar{\bar{\mathbf{I}}} \right] \cdot \mathbf{E}, \qquad l = 1,...,N. \qquad (1)$$

Here $\bar{\bar{\varepsilon}}_{\text{eff},l}$ is the effective dielectric function of each sub-array of wires, $\bar{\bar{\mathbf{I}}}$ is the identity matrix, $\varepsilon_h$ is the relative permittivity of the host medium and $\varepsilon_0$ is the vacuum permittivity. For conciseness, we identify the sub-array corresponding to the straight wires as sub-array *A*, while the racemic array of helical-shaped wires is identified as sub-array *B*. To ensure that the influence of each array on the others can be regarded as a macroscopic excitation, so that their near-field interaction is minimized, we consider a unit cell wherein the wires are placed as far apart as possible. We consider that the straight wire is placed in the center of the unit cell whereas the helical-shaped wires are placed close to the corners of the unit cell, as shown in Fig. 1. Moreover, as shown in Fig. 1 we consider that both arrays are oriented along the *z*-direction.

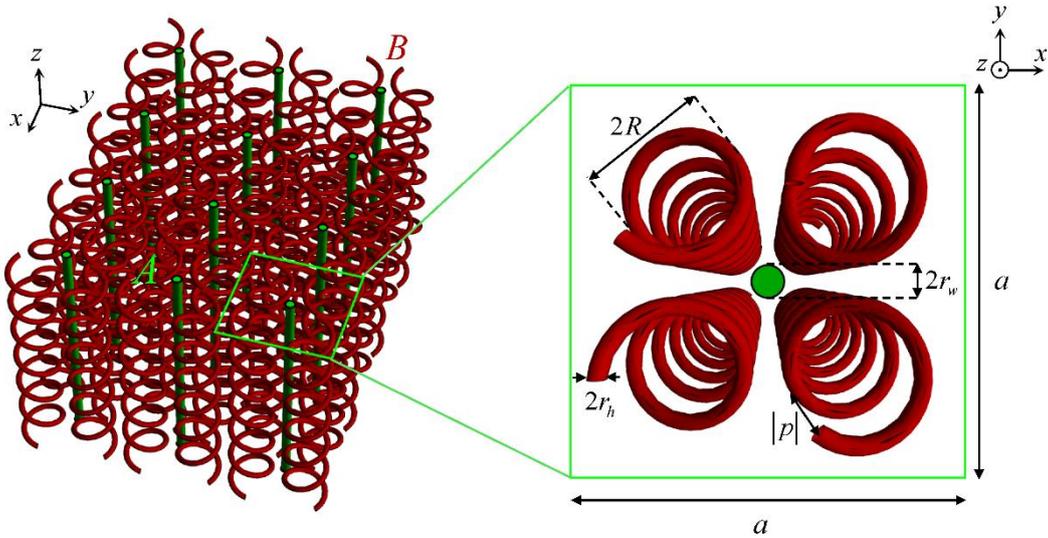

Fig. 1 (color online) Geometry of the wire metamaterial composed of two sub-arrays of metallic wires arranged in a periodic square lattice with period $a$: a standard wire medium formed by straight wires ( green wires denoted by sub-array *A*) with radius $r_w$, and a racemic array of helical-shaped wires (purple wires denoted by sub-array *B*) with helices with radius $R$ and pitch $|p|$, formed by wires with radius $r_h$. The wire arrays are oriented along the *z*-direction and embedded in a dielectric host with relative permittivity $\varepsilon_h$.

To begin with we determine the effective permittivity of the nested wire metamaterial. The effective dielectric function $\bar{\bar{\varepsilon}}_{\text{eff,A+B}}\left(\omega, -i\frac{d}{dz}\right)$ describes the response of the total polarization vector to the macroscopic electric field, which in our case is given by $\mathbf{P} = \mathbf{P}_A + \mathbf{P}_B$. The dependence of the effective permittivity on the wave vector $k_z \leftrightarrow -i\,d/dz$ accounts for the spatially dispersive (non-local) response of the wire arrays [4-5, 36-37]. Hence, it follows that the effective permittivity of the bulk metamaterial $\bar{\bar{\varepsilon}}_{\text{eff,A+B}}\left(\omega, -i\frac{d}{dz}\right)$ must be equal to:

$$\bar{\bar{\varepsilon}}_{\text{eff,A+B}} = \bar{\bar{\varepsilon}}_{\text{eff,A}} + \bar{\bar{\varepsilon}}_{\text{eff,B}} - \varepsilon_h \varepsilon_0 \bar{\bar{\mathbf{I}}}, \tag{2}$$

In what follows, we briefly review the effective medium response of each sub-array of wires. The standard wire medium configuration, which corresponds to the sub-array of wires *A*, consists of straight metallic wires with radius $r_w$ that are oriented parallel to each other and arranged in a periodic square lattice with period *a* [see Fig. 1]. To simplify our analysis, in this work we assume that the wires can be represented as perfect electric conductors (PEC). This approximation holds well when the radius of the rods is several times greater than the skin depth of the metal [59]. When the wires are made of a PEC, the effective permittivity tensor that characterizes the response of the bulk medium may be written as [4-5]:

$$\bar{\bar{\varepsilon}}_{\text{eff,A}}\left(\omega, -i\frac{d}{dz}\right) = \varepsilon_0 \varepsilon_h \left(\hat{\mathbf{u}}_x \hat{\mathbf{u}}_x + \hat{\mathbf{u}}_y \hat{\mathbf{u}}_y + \varepsilon_{zz,A} \hat{\mathbf{u}}_z \hat{\mathbf{u}}_z\right), \tag{3}$$

where $\varepsilon_{zz,A} = 1 - \dfrac{k_p^2}{k_0^2 \varepsilon_h - k_z^2}$ is the permittivity of the homogenized medium along the direction of the wires and $k_0 = \omega/c$ is the free space wave number, with $\omega$ being the angular frequency and $c$ the light speed in free space. Moreover, the operation $\hat{\mathbf{u}}_i \hat{\mathbf{u}}_i$, with $i = x, y, z$, stands for the tensor (outer) product of the two vectors. The parameter $k_p$ may be understood as the plasma wave number of the effective medium which depends on the geometry of the structure. Within a thin-wire approximation [23]

$$\left(k_p a\right)^2 = 2\pi \left[ \ln\left(\frac{a^2}{4 r_w (a - r_w)}\right) \right]^{-1}.$$

On the other hand, the sub-array of wires *B* consists of a racemic array of helical-shaped metallic wires periodic along the *x*- and *y*- directions and infinitely long along *z*, as illustrated in Fig. 1. The wires are arranged in a square unit cell that contains four helical-shaped wires: two right-handed helices and two left-handed helices, arranged in a checkerboard pattern. This sort of arrangement ensures that there are no bianisotropic effects in the response of the material. As a result, the spatially dispersive effective permittivity of the bulk metamaterial may be written as [36-37]:

$$\overline{\overline{\varepsilon}}_{\text{eff,B}}\left(\omega, -i\frac{d}{dz}\right) = \varepsilon_h \varepsilon_0 \left(\varepsilon_t \hat{\mathbf{u}}_x \hat{\mathbf{u}}_x + \varepsilon_t \hat{\mathbf{u}}_y \hat{\mathbf{u}}_y + \varepsilon_{zz,B} \hat{\mathbf{u}}_z \hat{\mathbf{u}}_z\right), \tag{4}$$

with

$$\varepsilon_{zz,B} = 1 - \dfrac{1}{\dfrac{k_0^2 \varepsilon_h}{k_{p1}^2} - \dfrac{k_z^2}{k_{p2}^2}} + \dfrac{A^2 k_0^2 \varepsilon_h}{\left(1 + \dfrac{A^2 k_0^2 \varepsilon_h}{\dfrac{k_0^2 \varepsilon_h}{k_{p1}^2} - \dfrac{k_z^2}{k_{p2}^2}}\right)\left(\dfrac{k_0^2 \varepsilon_h}{k_{p1}^2} - \dfrac{k_z^2}{k_{p2}^2}\right)^2}, \tag{5}$$

$$\text{and } \varepsilon_t = 1 + \frac{(2\pi R)^2}{V_{cell}} \frac{1}{C_1}. \tag{6}$$

Here $A = \pi R^2/p$, for helical-shaped wires with radius $R$ and pitch $p$ [see Fig. 1], and $V_{cell} = a^2 |p|$ is the volume of the unit cell. The parameters $k_{p1}$ and $k_{p2}$ are wave number parameters given by $k_{p1} = 4\pi \sqrt{\frac{p^2}{C_0 p^2 V_{cell} + 8C_1 \pi^2 R^2 V_{cell}}}$ and $k_{p2} = 4\pi \sqrt{\frac{1}{C_0 V_{cell}}}$, that only depend on the specific geometry of the helical-shaped wire array. Moreover, $C_0$ and $C_1$ are geometrical parameters with units of $\text{m}^{-1}$ whose definitions can be found in [60].

Thus, the effective permittivity of the proposed wire metamaterial is obtained inserting the effective permittivity of each sub-array of wires, given by Eqs. (3) and (4), into Eq. (2). It is important to mention that the effective medium model of the individual wire arrays is only valid under a thin wire approximation, so that $r_w \ll a$ and $r_h \ll a$, provided $a \ll \lambda$, with $\lambda$ the wavelength of operation and $k_\parallel a \ll \pi$, with $\mathbf{k} = k_z \hat{\mathbf{z}} + \mathbf{k}_\parallel$ [4-5].

As shown in Appendix A, the magnetic response of the nested wire metamaterial is that of the helical-shaped wire medium, which was thoroughly discussed in [36-37]. Such magnetic response plays no role in the emergence of the Fano resonances discussed ahead in this article. Hence, we will focus our study on the electric response of the nested wire metamaterial.

Within an effective medium approach, the photonic modes that can propagate in the structure can be calculated from the characteristic equation of the problem using the effective parameters of the metamaterial. This equation can be obtained replacing the effective permittivity tensor of the metamaterial into the Maxwell equations and calculating the plane-wave solutions for a spatial variation of the form $e^{i\mathbf{k}\cdot\mathbf{r}}$, so that $\nabla = i\mathbf{k}$, with $\mathbf{k} = \mathbf{k}_t + k_z \hat{\mathbf{z}}$ and $\mathbf{k}_t = k_x \hat{\mathbf{x}} + k_y \hat{\mathbf{y}}$ the transverse component of the wavevector. For propagation in the *xoz*-plane, so that $\partial/\partial y = 0$ and $k_y = 0$, it is easily checked that the dispersion of the Transverse Magnetic (TM) eigenwaves (with $\mathbf{H} = H_y \hat{\mathbf{y}}$) can be calculated from the solution of the characteristic equation:

$$\det\left[\left(k^2\left(\bar{\bar{\mathbf{I}}} - \hat{\mathbf{u}}_y\hat{\mathbf{u}}_y\right) - \mathbf{k}\mathbf{k}\right) \cdot \left(\frac{\bar{\bar{\varepsilon}}_{\text{eff,A+B}}}{\varepsilon_0}\right)^{-1} - \bar{\bar{\mathbf{I}}}(\omega/c)^2\right] = 0, \quad (7)$$

with $k^2 = |\mathbf{k}|^2$ and $\mathbf{k}\mathbf{k}$ is the tensor product operation of $\mathbf{k}$ with itself. For a given value of $\mathbf{k}_t$ the dispersion relation (7) corresponds to a cubic equation on $k_z^2$, thus yielding three different solutions (eigenmodes) with positive frequency $\omega$ and three different solutions with negative (and symmetric) frequency $\omega$.

Let us consider a metamaterial structure wherein the wires in sub-array *A* have radius $r_w/a = 0.005$, while the wire medium forming sub-array *B* is characterized by helices with radius $R = 0.1a$, pitch $|p| = 0.1a$, and wires with radius $r_h/a = 0.005$. For this set of structural parameters $C_0 a \approx 38.28$ and $C_1 a \approx 19.5$. Moreover, both sets of wires are embedded in a vacuum. The band diagram of this metamaterial structure, calculated with the numerical simulator [61] and the effective medium model, when $\mathbf{k}_t = 0$, is shown in Fig. 2.

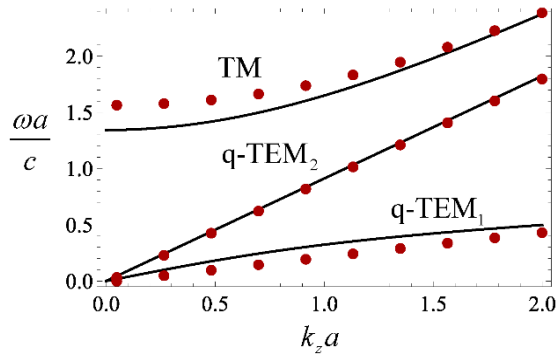

Fig. 2 – (color online) Band diagram: $\omega a/c$ of the nested wire metamaterial structure calculated as a function of the normalized wave vector $k_z a$ when $\mathbf{k}_t = 0$. The sub-array *A* has straight wires with $r_w/a = 0.005$. The helices in sub-array $B$ have $R = 0.1a$, $|p| = 0.1a$, $r_h/a = 0.005$, $C_0 a \approx 38.28$ and $C_1 a \approx 19.5$. For a fixed value $k_z a$, the wire metamaterial supports 3 eigenwaves with positive $\omega$: two quasi-Transverse Electromagnetic waves, denoted by q-TEM$_1$ and q-TEM$_2$, and a TM mode. The solid black lines correspond to the results calculated with the effective medium theory and the discrete red symbols to the results obtained from full-wave simulations [61].

The results obtained with the homogenization model are in good agreement with the full-wave results and reveal that the metamaterial supports three eigenmodes: two quasi-Transverse Electromagnetic (q-TEM) waves, with a nearly linear dispersion, and a TM mode, similar to what is observed in the original nested wire medium configuration [39]. We designate these modes by q-TEM$_1$, q-TEM$_2$ and TM. Each q-TEM mode is associated with a different sub-array of wires [5, 23, 29, 39]. Specifically, one quasi-TEM mode is associated with a field localized close to helical-shaped wires (q-TEM$_1$), while the other mode (q-TEM$_2$) corresponds to a field concentrated near the straight wires.

## III. Fano resonances in the wire metamaterial

Here we apply the effective medium model of the nested wire metamaterial that was developed in the previous section to characterize the scattering of electromagnetic waves in a metamaterial slab with thickness $h$. We consider TM-polarized incident waves propagating in the *xoz* plane (i.e., $k_y = 0$ and magnetic field polarized along the *y*-direction) with an angle of incidence $\theta_{inc}$. A TM-polarized incident wave can excite plane waves in the metamaterial slab with transverse wave vector $\mathbf{k}_t = k_x \hat{\mathbf{x}}$, with $k_x = (\omega/c) \sin(\theta_{inc})$. A representative system geometry is depicted in Fig. 3.

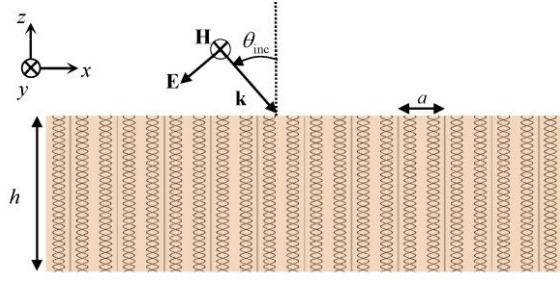

Fig. 3 – (color online) Geometry of the scattering problem under study: a wire metamaterial slab with thickness $h$ is illuminated by a TM-polarized plane wave with an incidence angle $\theta_{\text{inc}}$. The metamaterial slab has period $a$ and is composed by two sub-arrays of wires: sub-array A formed by straight metallic wires and sub-array B formed by a racemic helical-shaped wire medium.

For the studied configuration, the relevant components of the electromagnetic field are $H_y$, $E_x$ and $E_z$. We use a plane wave expansion of the fields and write them as a superposition of exponentials propagating in the *xoz* plane. The magnetic field distribution in all space may be written as:

$$H_y(z,\omega) = e^{ik_x x} \frac{E^{inc}}{\eta_0} \begin{cases} e^{\gamma_0 z} - \rho e^{-\gamma_0 z} & z > 0 \\ A_1^+ e^{\gamma_{\text{q-TEM}_1}(z+h)} + A_1^- e^{-\gamma_{\text{q-TEM}_1}(z+h)} + A_2^+ e^{\gamma_{\text{q-TEM}_2}(z+h)} \\ \quad + A_2^- e^{-\gamma_{\text{q-TEM}_2}(z+h)} + A_3^+ e^{\gamma_{\text{TM}}(z+h)} + A_3^- e^{-\gamma_{\text{TM}}(z+h)} & -h < z < 0 \\ T e^{\gamma_0(z+h)} & z < -h \end{cases} \quad (8)$$

Here $A_i^\pm$, with $i = 1, 2, 3$, correspond to the complex amplitudes of the counter-propagating waves inside the metamaterial slab, each pair associated with one eigenmode propagating in the slab, and $\rho$ and $T$ are the reflection and transmission coefficients, respectively. Moreover, the propagation constant in free space is determined by $\gamma_0 = \sqrt{k_x^2 - \omega^2 \mu_0 \varepsilon_0}$, while the propagation constants of each mode ($\gamma_{\text{q-TEM}_1}$, $\gamma_{\text{q-TEM}_2}$ and $\gamma_{\text{q-TM}}$) in the metamaterial, which are of the form $\gamma^2 = -k_z^2$, are calculated from the solutions of Eq. (7) with $\mathbf{k} = k_x \hat{\mathbf{x}} + k_z \hat{\mathbf{z}}$.

The electric field distribution can be calculated from the magnetic field in Eq. (8) using

$$\mathbf{E} = \frac{1}{-i\omega}\left[\overline{\overline{\varepsilon}}_{\text{eff,A+B}}\left(\omega, -i\frac{d}{dz}\right)\right]^{-1} \cdot \nabla \times \mathbf{H},$$ with the permittivity tensor given by Eq. (2), and

it is equal to

$$E_x(z,\omega) = \frac{e^{ik_x x} E^{inc}}{i\omega\varepsilon_0\eta_0} \begin{cases} \gamma_0\left(e^{\gamma_0 z} + \rho e^{-\gamma_0 z}\right) & z > 0, \\ \frac{1}{\varepsilon_h \varepsilon_t}\gamma_{\text{q-TEM}_1}\left(A_1^+ e^{\gamma_{\text{q-TEM}_1}(z+h)} - A_1^- e^{-\gamma_{\text{q-TEM}_1}(z+h)}\right) & \\ +\frac{1}{\varepsilon_h \varepsilon_t}\gamma_{\text{q-TEM}_2}\left(A_2^+ e^{\gamma_{\text{q-TEM}_2}(z+h)} - A_2^- e^{-\gamma_{\text{q-TEM}_2}(z+h)}\right) & -h < z < 0 \\ +\frac{1}{\varepsilon_h \varepsilon_t}\gamma_{\text{TM}}\left(A_3^+ e^{\gamma_{\text{TM}}(z+h)} - A_3^- e^{-\gamma_{\text{TM}}(z+h)}\right) & \\ \gamma_0 T e^{\gamma_0(z+h)} & z < -h \end{cases} \quad (9)$$

$$E_z(z,\omega) = -\frac{e^{ik_x x} E^{inc} k_x}{\omega\eta_0} \begin{cases} e^{\gamma_0 z} - \rho e^{-\gamma_0 z} & z > 0 \\ \frac{1}{\varepsilon_{\text{eff,A+B},zz}\left(\omega, i\gamma_{\text{q-TEM}_1}\right)}\left(A_1^+ e^{\gamma_{\text{q-TEM}_1}(z+h)} + A_1^- e^{-\gamma_{\text{q-TEM}_1}(z+h)}\right) & \\ +\frac{1}{\varepsilon_{\text{eff,A+B},zz}\left(\omega, i\gamma_{\text{q-TEM}_2}\right)}\left(A_2^+ e^{\gamma_{\text{q-TEM}_2}(z+h)} + A_2^- e^{-\gamma_{\text{q-TEM}_2}(z+h)}\right) & -h < z < 0 \\ +\frac{1}{\varepsilon_{\text{eff,A+B},zz}\left(\omega, i\gamma_{\text{TM}}\right)}\left(A_3^+ e^{\gamma_{\text{TM}}(z+h)} + A_3^- e^{-\gamma_{\text{TM}}(z+h)}\right) & \\ T e^{\gamma_0(z+h)} & z < -h \end{cases} \quad (10)$$

To characterize the scattering properties of the metamaterial slab, i.e. to calculate $\rho$ and $T$, as well as the complex amplitudes of the modes inside the slab $A_i^\pm$, one can impose suitable boundary conditions at the metamaterial-air interfaces. The classical boundary conditions describe the behavior of the tangential components of the electric and magnetic fields at the interfaces. Since there are no surface currents or surface magnetization at the interfaces, the tangential components of the electric and magnetic fields are continuous at $z = 0, -h$, so that:

$$\lfloor E_x \rfloor_{z=0,-h} = 0 \qquad (11a)$$

$$\left\lfloor H_y \right\rfloor_{z=0,-h} = 0 \tag{11b}$$

where $\left\lfloor F \right\rfloor_{z=z_0} = F_{z=z_0^+} - F_{z=z_0^-}$, i.e. $\left\lfloor \; \right\rfloor$ is an operator that evaluates the discontinuity of $F$ at a pertinent interface $z = z_0$. However, because of the spatial dispersion of the wire arrays, applying solely the classical boundary conditions to solve this scattering problem results in an underdetermined system of equations. To overcome this problem, one must also specify additional boundary conditions at the interfaces [24, 62-65]. As shown in [23, 39] these boundary conditions characterize the behavior of the microscopic currents flowing in each set of wires at the interfaces. Considering that both sub-arrays are severed at $z = 0, -h$, the current flowing in the wires should vanish at these interfaces. Since the interaction of each array with the other can be viewed as a macroscopic excitation, the contribution of the polarization vector of each sub-array to the total polarization vector ( $\mathbf{P} = \mathbf{P}_A + \mathbf{P}_B$ ) is related to the current flowing through its wires as $\mathbf{P}_l \cdot \hat{\mathbf{z}} = \dfrac{1}{-i\omega} \dfrac{I_l}{a^2}$, with $l = A, B$. Hence, stating that the current in each set of wires vanishes at the interface is equivalent to enforcing that [39]:

$$\left. \mathbf{P}_l \cdot \hat{\mathbf{z}} \right|_{z=0,-h} = 0, \text{ with } l = A, B. \tag{12}$$

It is easily checked from Eq. (1) that Eq. (12) may also be written using the normal component of the electric field, so that the additional boundary conditions are:

$$\left( \varepsilon_{zz,A}\left( \omega, -i\frac{d}{dz} \right) - 1 \right) E_z \bigg|_{z=0,-h} = 0 \tag{13a}$$

$$\left( \varepsilon_{zz,B}\left( \omega, -i\frac{d}{dz} \right) - 1 \right) E_z \bigg|_{z=0,-h} = 0 \tag{13b}$$

Using Eqs. 11a-b and 13a-b allow us to obtain a linear system of equations that can be numerically solved to determine the scattering parameters of the metamaterial slab. In what follows, we characterize the scattering properties of a metamaterial slab with thickness $h = 5a$ for incident TM-waves with an incidence angle $\theta_{inc} = 60°$. The sub-arrays of wires have the same structural parameters as considered in the calculations depicted in Fig. 2. The transmission properties of the slab, calculated as a function of the operation frequency, are shown in Fig. 4a-b where we compare the results obtained with the homogenization model (blue solid curves) with the numerical simulations (green dashed curves) obtained using the full wave simulator [61].

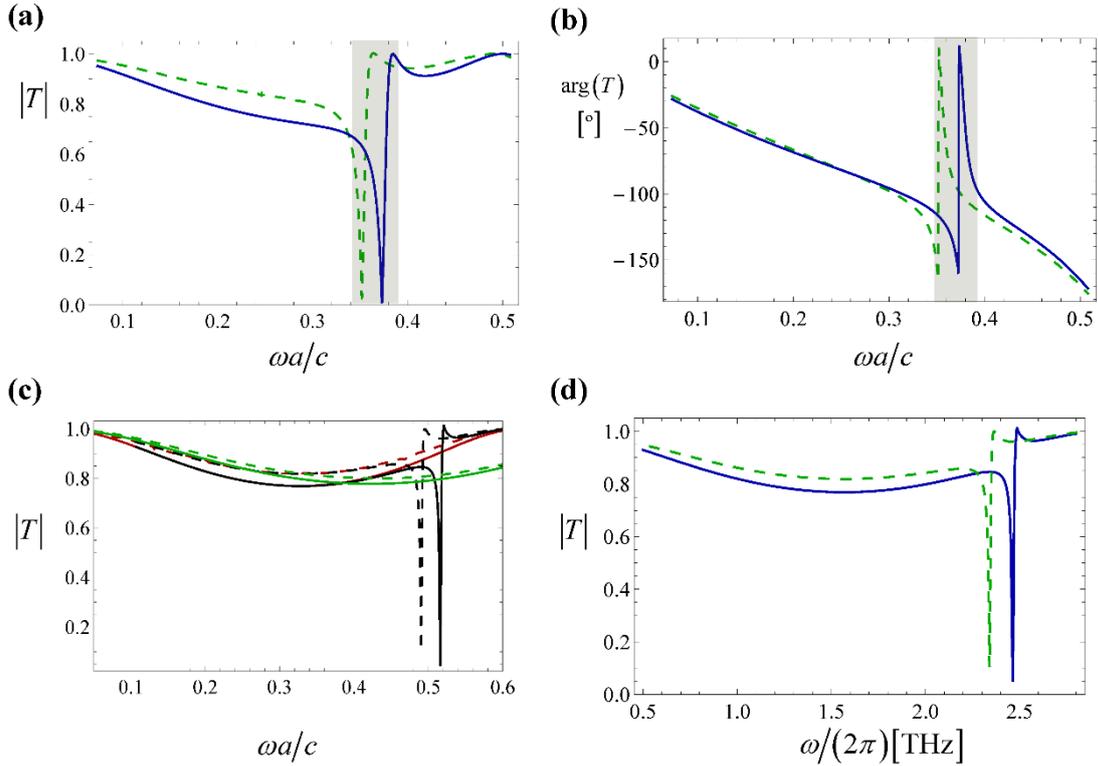

Fig. 4 –(color online) **a)** Amplitude and **b)** phase of the transmission coefficient calculated using the effective medium model (blue solid curve) and the full wave simulations (green dashed curve) as a function of the normalized frequency $\omega a/c$ for a metamaterial slab with thickness $h = 5a$. The remaining structural parameters are the same as considered in Fig. 2. **c)** Amplitude of the transmission coefficient of the nested wire metamaterial slab (black curves), helical-shaped wire metamaterial slab (red curves) and standard wire medium slab (green curves) as a function of the normalized frequency $\omega a/c$. The pitch of the helices is 2.5 times larger than in the configuration used in panel **a)**, i.e. $|p| = 0.25a$, so that $C_0 a \approx 29.42$ and $C_1 a \approx 14.72$ while the remaining of the parameters are as in panel a). **d)** Amplitude of the transmission coefficient calculated as a function of frequency for a metamaterial slab with thickness $h = 5a$ and $a = 10\,\mu m$,

considering aluminum wires (full-wave numerical results) and PEC wires (theoretical results). The remaining structural parameters are the same as in **c)**. In all panels the TM incident waves illuminate the slab at an incidence angle $\theta_{inc} = 60º$. Moreover, the effective medium model results correspond to the solid curves, whereas the dashed curves depict the full-wave simulation results.

As seen, apart from a small frequency shift, the results calculated using the effective medium model and the numerical simulations have a good agreement, indicating that the homogenization formalism can accurately characterize the electromagnetic response of the nested wire metamaterial. It is important to mention that we numerically checked (not shown) that when the helices have radii larger than $R \geq 0.2a$, the effective medium model fails to accurately characterize the electromagnetic response of the metamaterial. This happens because for helices with large radii there is a non-negligible near-field coupling between each sub-array of wires.

Quite interestingly, the transmission characteristic of the slab reveals the emergence of a sharp resonance with asymmetric line shape around the normalized frequency $\omega a/c \approx 0.36$, highlighted in Fig. 4a-b as a shaded gray region. The asymmetric shape of this resonance is very distinct from a Lorentzian-type resonance and can clearly be identified as a characteristic feature of a Fano resonance [40,41]. Notably, increasing the pitch of the helical-shaped wires induces a small blue shift in the Fano resonance [compare Figs. 4a and 4c]. Additionally, it can be checked that the resonance linewidth narrows when the pitch increases.

In Ref. [39], we showed that the unit cell of the nested wire metamaterial must have some sort of structural asymmetry for the Fano resonances to emerge in the medium, either by considering wires formed by different plasmonic materials, or by terminating each sub-array of wires differently at the interfaces. In this study, we demonstrate the presence of Fano-type resonances in the scattering properties of a nested wire metamaterial even when the wires are terminated uniformly and are made of the same material. The structural asymmetry is achieved by incorporating two sub-arrays of wires with distinct shape.

We also examined the scattering of incident TM-polarized waves in nested wire metamaterials where all helical shaped wires have the same handedness. This medium exhibits magnetoelectric (chiral) coupling [36, 60], allowing some of the energy from the incident TM wave to couple with the TE mode. Our analysis revealed that while this structure may also exhibit Fano resonances for TM-incidence, the resonance linewidth is notably broader compared to the original configuration. Such effect ultimately constrains the practical applications of the resonant behavior for a TM-incident polarization.

To demonstrate in a conclusive manner that the emergence of the Fano resonance is linked to the interaction between both sub-arrays of wires, we also depict in Fig. 4c the response of each individual sub-array of wires alone. Under the effective medium approach, the scattering properties of each sub-array of wires are calculated neglecting the contribution of the other sub-array in the effective response of the nested metamaterial. As we can see, the response of the nested wire metamaterial is totally different from that of each sub-array of wires. Particularly, around the Fano resonance both the transmission characteristic of sub-array *A* (straight wires), depicted as green curves in Fig. 4c, and that of sub-array *B* (helical wire medium), depicted as red curves in Fig. 4c, both exhibit broad weak dipole resonances (centered around $\omega a/c \approx 0.32$ and $\omega a/c \approx 0.42$). Generically a Fano resonance emerges from the interference between two resonances, one with a narrow lineshape and another with a broad lineshape [52]. These resonances may originate from the interaction between different resonators [39] or from the interaction of different modes of the same resonator [52, 66]. The emergence of Fano resonances in nested wire metamaterials is related to the macroscopic interaction that takes place when combining both sets of wires within the same unit cell. This interaction induces a phase disparity in the currents flowing through the two wire arrays, thereby generating a distinct narrow antibonding mode that is associated with a quadrupole/magnetic resonance [39]. Indeed, at the peak of transmission of the Fano resonance the subwavelength wires, which to a first approximation can be regarded as dipoles, have dipole moments with the same amplitude but oscillate out of phase, so that the net dipole moment is zero and an incoming wave is completely transmitted. Similar effects may be observed in other types of metamaterials [68]. In summary, it is the interference between this narrow quadrupole/magnetic resonance and the broad dipolar resonance of each set of wires that gives rise to the Fano asymmetric shape in the nested wire metamaterial.

To better illustrate the effect, we calculated the $z$-component of the polarization vector of each sub-array of wires ($\mathbf{P}_i \cdot \hat{\mathbf{z}} = \frac{1}{-i\omega} \frac{I_i}{a^2}$) using the proposed effective medium formalism, at the frequency corresponding to the peak of transmission at the Fano resonance, which for the metamaterial slab considered in Fig. 4c corresponds to the frequency $\omega a/c \approx 0.5208$. The results are shown in Fig. 5a and demonstrate that at the Fano resonance the $z$-component of the polarization vectors in each set of wires have nearly identical amplitude and are in opposition of phase, so that the total polarization vector along the $z$-direction vanishes. This is substantiated in Fig. 5b where we show the full-wave simulation results of the microscopic current in the wires at the frequency corresponding to the peak of transmission ($|T| \approx 1$) calculated with the simulator [61]. It is important to emphasize that while the microscopic current in the helical-shaped wires flows along the helical path, the spatial-averaged current is oriented parallel to the straight wires ($z$-direction). From Fig. 5b we see that the counterclockwise microscopic flow of the current in the helices originates an average current that is directed in the opposite direction of the current flow in the straight wires. Hence, consistent with the homogenization results, the numerical simulations reveal that the physical reason behind the peak of transmission at the Fano resonance is the opposite signed polarization vectors in each sub-array of wires.

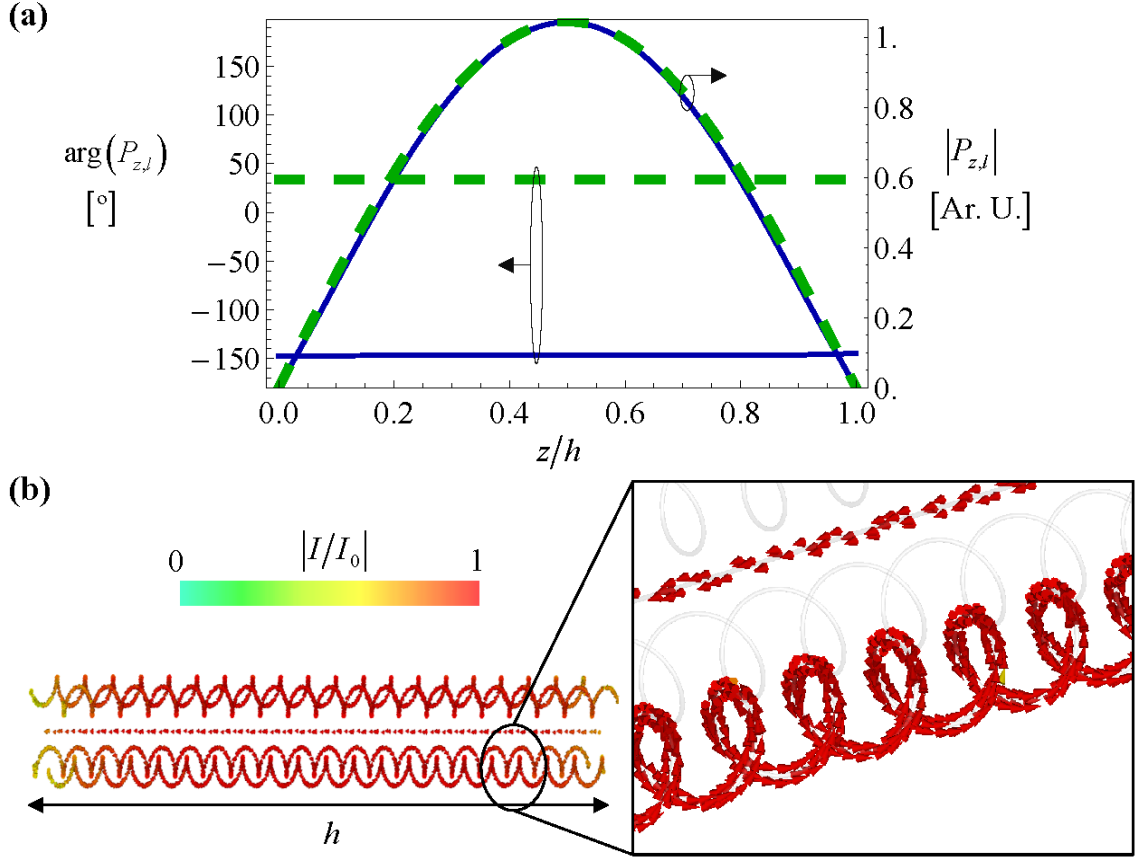

Fig. 5 -(color online) **a)** Normalized amplitude and phase of the *z*-component of the polarization vector of each sub-array of wires at the frequency of operation $\omega a/c \approx 0.5208$ (corresponding to the transmission peak at the Fano resonance). The green curves correspond to the polarization vector of sub-array *A* (array of straight wires), whereas the blue curves are the results calculated for sub-array *B* (racemic array of helical-shaped wires). **b)** Normalized amplitude of the microscopic current density distribution in a unit cell of the wire metamaterial slab calculated at the transmission peak of the Fano resonance. In both panels the geometric parameters are the same as those considered in Fig. 4c.

To have a better understanding of the robustness of the emergence of these resonances in our system, in what follows we determine the Fano resonance frequency for some variations of the structural parameters. Importantly, as the emergence of a Fano resonance in the metamaterial response is linked to the formation of a sub-radiant mode which results from a null net polarization vector, one can determine the peak transmission frequency at the Fano resonance simply by calculating the value of the net polarization vector $\mathbf{P} \cdot \hat{\mathbf{z}} = (\mathbf{P}_A + \mathbf{P}_B) \cdot \hat{\mathbf{z}}$. In Fig. 6a we show the net polarization vector calculated as a function of the frequency of operation and incidence angle for the same metamaterial parameters as considered in Fig. 4c. The result reveals that the frequency of the transmission peak at the Fano resonance ($\omega a/c \approx 0.5208$), which corresponds to the null of the net polarization vector (dark region in Fig. 6a), is highly insensitive to variations of the incidence angle. Therefore, for $\omega a/c \approx 0.5208$ and a wide range of incidence angles the currents excited in the sub-arrays of wires have the same amplitude and flow in opposite directions. The physical reason behind this response can be inferred from the band diagram shown in Fig. 2. For small frequencies the metamaterial only supports two bulk modes, the q-TEM modes, each one associated with one set of wires. We checked that for small frequencies the isofrequency contours of the quasi-TEM modes are nearly flat, i.e. are vastly insensitive to variations of $k_x = \sin(\theta_{\text{inc}})\omega/c$.

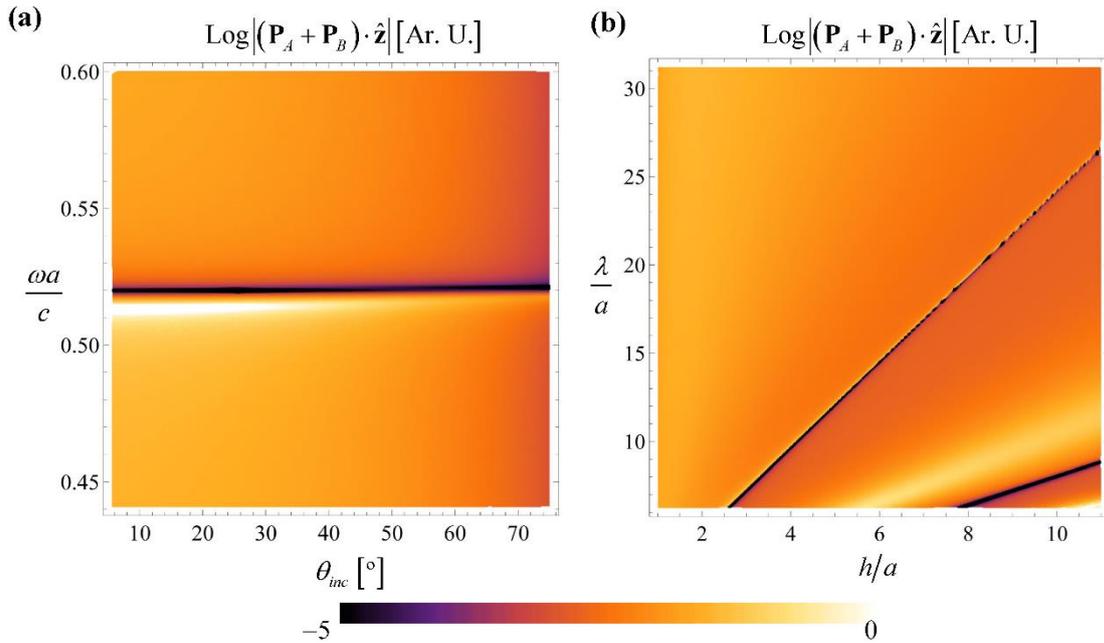

Fig. 6 (color online)  **a)** Amplitude of the net polarization vector $\left|\left(\mathbf{P}_A + \mathbf{P}_B\right) \cdot \hat{\mathbf{z}}\right|$ (in arbitrary logarithmic units) as a function of the frequency of operation and incidence angle for a metamaterial slab with the same parameters considered in Fig. 4c **b)** Similar to **a)** but we fix the incidence angle $\theta_{inc} = 60°$ and vary the wavelength of operation $\lambda = 2\pi c/\omega$ and the thickness of the slab. In both panels the dark color points (null net polarization vector) correspond to Fano resonances.

Additionally, we performed another parametric study where we fixed the incidence angle at $\theta_{inc} = 60°$ and varied the thickness of the metamaterial slab. The corresponding results are shown in Fig. 6b and confirm that a Fano resonance appears for all the considered values of *h*, revealing also the emergence of other (higher order) Fano resonance for larger frequencies (smaller wavelengths of operation). Furthermore, it is seen that the Fano resonance wavelength is proportional to the thickness of the slab, similar to the resonance wavelength of electric dipole antennas. Thus, varying the length of the slab can provide for a tuning mechanism to obtain a Fano resonance in a desired frequency range. Since the wire arrays are made from the same metallic material, provided that its conductivity is high enough, the proposed metamaterial can be effectively scaled to operate across a broad frequency range, spanning from microwave regime up to THz frequencies.

To demonstrate that the Fano resonance can be excited at THz frequency, we consider that both sets of wires are made from aluminum and determine the scattering properties of the metamaterial slab with the full-wave simulator for a lattice period of $a = 10 \, \mu m$ while the remaining structural parameters are as in Fig. 4c. The aluminum material is modeled in the simulator by a conductivity $\sigma = 3.5607 \times 10^7 \, \text{S/m}$. The results are shown in Fig. 4d and reveal the emergence of a Fano sharp resonance near $\omega/(2\pi) \approx 2.37 \text{THz}$, thus indicating that the proposed metamaterial can indeed be designed to operate in the THz regime. Moreover, in Fig 4d we also overlap the homogenization model results calculated for PEC wires. Interestingly, it is seen that the frequency shift between both results is very similar to that in Fig 4c, indicating that the aluminum wires remain good conductors at this frequency range.

Finally, we also considered the possibility of fixing the incidence angle and varying the pitch of the helical-shaped wires. The results for a set of structural parameters as in Fig 4c are shown in Fig. 7a and reveal that as the pitch increases, the Fano resonances becomes increasingly narrower (the dark region, wherein $\mathbf{P} \cdot \hat{\mathbf{z}} \approx 0$, becomes increasingly thinner as the pitch of the helices increases), in line with the results of Fig. 4a and Fig. 4c. A more precise quantification of the resonance's linewidth can be inferred from the frequency bandwidth $\Delta \omega_L$ spanning from low to high transmission of the Fano resonance [71] (see the inset in Fig. 7b). This bandwidth is normalized to the peak transmission frequency at the Fano resonance $\omega_{T=1}$. The normalized frequency bandwidth from low to high transmission frequencies for the Fano resonance, calculated across various pitch values $p$ of the helices using the same structural parameters as depicted in Fig. 4c, is presented in Fig. 7b. These results corroborate that as the pitch $p$ increases, the resonance's linewidth diminishes, consistent with the findings of Fig. 7a.

Since the linewidth of a resonance is inversely proportional to its quality ($Q$) factor, our results indicate that as the pitch of the helices increases, the $Q$-factor also increases. In Appendix C, we provide a description of the method for determining the $Q$-factor of a Fano resonance in the metamaterial slab. We calculated the $Q$-factor for a metamaterial slab as a function of the pitch of the helices, for the same structural parameters as considered in Fig. 4c. These results are depicted in Fig. 7c revealing that the $Q$-factor can exceed $Q > 8000$ when $p/a > 0.75$. We also checked (not shown) that the $Q$-factor does not vary significantly with the transverse component of the wave-vector, which goes in line with the results shown in Fig. 6a, wherein the linewidth of the region wherein $\mathbf{P} \cdot \hat{\mathbf{z}} \approx 0$ is almost independent of the incidence angle. It is important to mention that the *Q*-factor of the Fano resonance may be set *arbitrarily* large by increasing the pitch of the helices. We also find relevant to mention that the enhancement of the *Q*-factor depicted in Fig. 7c may also be seen as a signature of a bound state in the continuum (BIC). A sharp Fano resonance in the scattering parameters is often an indicator of a trapped state [72, 73]. Indeed, if the linewidth of the resonance vanishes, the radiative lifetime becomes infinite, which is consistent with a BIC. In our configuration, the linewidth of the resonance is tied to the degree of bulk structural asymmetry in the metamaterial, specifically associated with the pitch of the helices. Breaking the symmetry ($p \neq \infty$) induces radiative coupling, yielding a Fano resonance with finite linewidth connected with a quasi-bound state in the continuum (q-BIC) [74-77]. However, in the limit when $p \to \infty$ there are no Fano resonances. The physical reason behind such response is that when $p \to \infty$ the metamaterial becomes a straight wire medium which, as shown in Fig. 4c, does not exhibit any Fano resonance. It was shown in [39] that even when both sets of straight wires have different radii, the metamaterial lacks the capability to exhibit Fano resonances when both wire sub-sets are composed of the same material. This suggests that the level of asymmetry in such structure is insufficient to induce Fano resonances.

The connection between the degree of structural asymmetry in the nested wire metamaterial and the emergence of the Fano resonance may also be inferred from the dimensionless Fano factor *q* that characterizes the spectral shape asymmetry of the Fano resonance. The lineshape of a Fano resonance can be generically described using the frequency dependent equation [40, 52]:

$$\sigma(\omega) = \frac{1}{1+q^2} \frac{\left(q\Gamma + 2(\omega-\omega_0)\right)^2}{\Gamma^2 + 2(\omega-\omega_0)^2}, \tag{14}$$

where $\omega_0$ and $\Gamma$ are the resonant frequency and width, respectively. To make the lineshape amplitude compatible with the transmission spectra, we normalized it to fall within the range of zero to one. Through a numerical fitting process, we matched the lineshape described in Eq. (14) with the transmission characteristics of the slab within the frequency range of the Fano resonance. This allowed us to determine both the Fano factor $q$ and the linewidth of the resonance $\Gamma$. We calculated $q$ and $\Gamma$ using the same metamaterial slab as illustrated in Fig. 7c. These calculations were performed as a function of the pitch of the helices, and the results are depicted in Fig. 7d. Notably, the Fano parameter and the linewidth of the resonance exhibit a significant reduction as $p$ increases. It is worth noting that as $q \to 0$ and $\Gamma \to 0$, in accordance with Eq. (14), the Fano resonance ceases to exist.

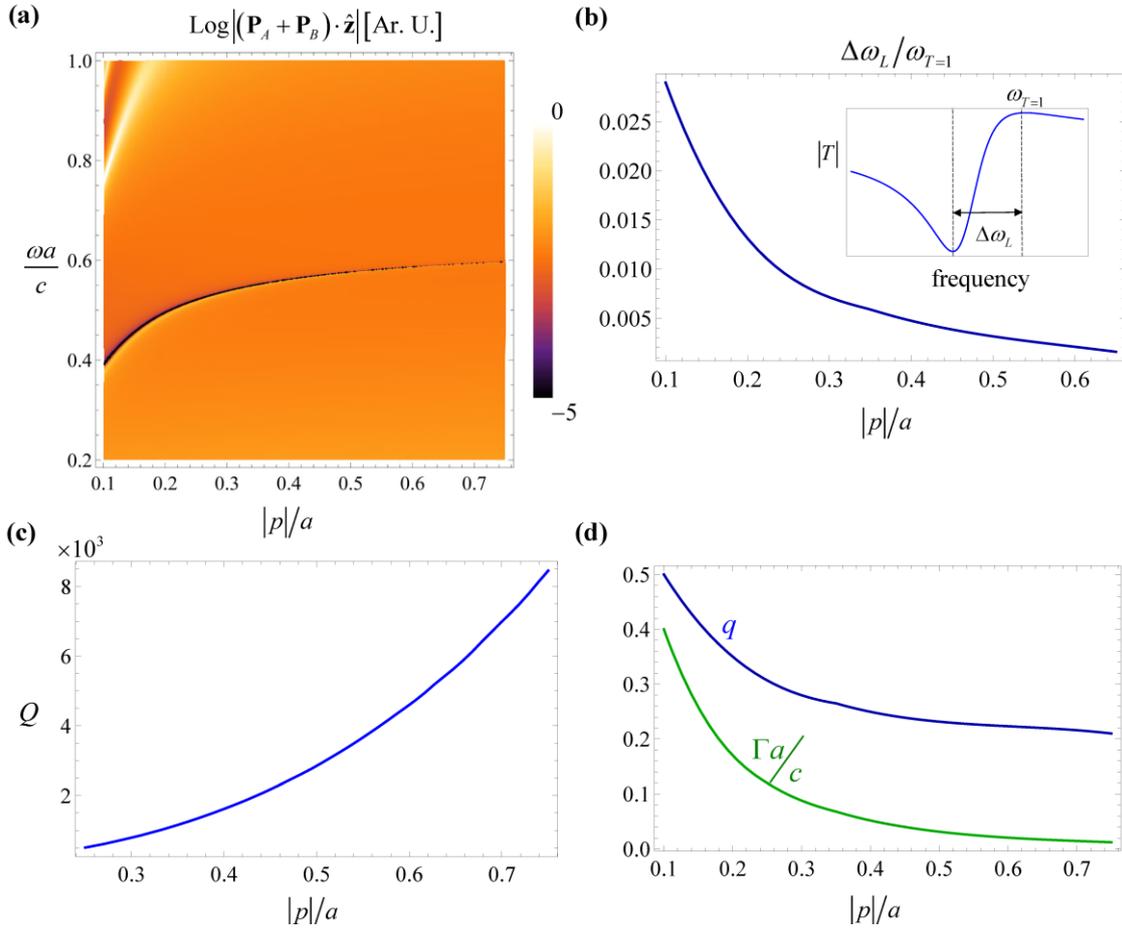

Fig. 7 - (color online) **a)** Amplitude of the net polarization vector $|(\mathbf{P}_A + \mathbf{P}_B) \cdot \hat{\mathbf{z}}|$ (in arbitrary logarithmic units) as a function of the frequency of operation and the pitch of the helical-shaped wires. The dark color points (null net polarization vector) correspond to Fano resonances. **b)** Frequency bandwidth spanning from low to high transmission $\Delta\omega_L$ of the Fano resonance normalized to the peak transmission frequency at the Fano resonance $\omega_{T=1}$. **c)** Quality ($Q$) factor of the Fano resonance. **d)** Fano parameter $q$ (blue curve) and

normalized width of the Fano resonance $\Gamma a/c$ (green curve). In panels **b)-d)** the metamaterial parameters are the same as those considered in Fig. 4c and the quantities are calculated as a function of the pitch of the helical-shaped wires.

## IV. Conclusions

In this work we studied the effective medium response of a nested wire metamaterial formed by a set of straight wires and a racemic array of helical-shaped wires. The proposed effective model was validated against full-wave simulations. We investigated the scattering of plane waves by a slab of such a metamaterial and showed that the coupling between each set of wires can give rise to sharp Fano resonances with asymmetric line shape and high quality (*Q*) factor. The physical origin of these resonances is related to the interaction between the field scattered by each set of wires when there is a counterpropagating flow of currents in each sub-array. Furthermore, we also demonstrated that by tuning the pitch of the helical shaped sub-wire array, it is possible to drastically enhance the Fano asymmetric factor and the *Q*-factor of the Fano resonance. We envision that the scattering properties of the proposed metamaterial, with sharp Fano-type variations between opaque and transparent states, can have interesting applications in sensing and switching devices, ranging from the microwave up to THz frequencies.

Acknowledgments:

This work was funded by Instituto de Telecomunicações (IT) under project HelicalMETA - UIDB/50008/2020. D. E. Fernandes acknowledges financial support by IT-Lisbon under the research contract with reference C-0042-22. S. Lannebère acknowledges financial support by IT-Coimbra under the research contract with reference DL 57/2016/CP1353/CT0001. T. A. Morgado acknowledges FCT for research financial support with reference CEECIND/04530/2017 under the CEEC Individual 2017, and IT-Coimbra for the contract as an assistant researcher with reference CT/No. 004/2019-F00069. The authors acknowledge the preliminary simulations performed by Francisco Faim regarding the validity of the homogenization model.

## Appendix A: The effective permeability function

In this section we derive the magnetic response of the nested wire metamaterial by extending the formalism described in [29, 39] to calculate its effective permeability.

It is easily checked that if both sub-arrays of wires interact with one another as macroscopic excitations their impact on the macroscopic magnetic polarization vector **J** can be written as follows:

$$\mathbf{J}_l = \left[ \bar{\bar{\mu}}_{\text{eff},l} - \mu_0 \bar{\bar{\mathbf{I}}} \right] \cdot \mathbf{H}, \qquad l = A, B. \tag{A1}$$

where $\bar{\bar{\mu}}_{\text{eff},l}$ is the effective permeability of each sub-array and $\mu_0$ is the vacuum permittivity. Since the standard wire medium configuration (sub-array $A$) does not have a magnetic response [9], i.e. $\mu_{\text{eff},A} = \mu_0$, the total magnetic polarization vector is $\mathbf{J} = \mathbf{J}_B$ and therefore the effective permeability of the proposed nested metamaterial is simply equal to the effective permeability of the helical-shaped wire medium, so that:

$$\bar{\bar{\mu}}_{\text{eff},A+B} = \bar{\bar{\mu}}_{\text{eff},B}. \tag{A2}$$

The helical-shaped wire medium has a strong magnetic response, behaving as a magnetic uniaxial wire medium [37]. Its effective permeability tensor [36, 37] may be written as:

$$\bar{\bar{\mu}}_{\text{eff},B}(\omega, k_z) = \mu_0 \left( \hat{\mathbf{u}}_x \hat{\mathbf{u}}_x + \hat{\mathbf{u}}_y \hat{\mathbf{u}}_y + \mu_{zz} \hat{\mathbf{u}}_z \hat{\mathbf{u}}_z \right), \tag{A3}$$

with $\mu_{zz} = \left(1 + \dfrac{A^2 k_0^2}{\dfrac{k_0^2}{k_{p1}^2} - \dfrac{k_z^2}{k_{p2}^2}}\right)^{-1}$.

## Appendix B: The dispersion equation

In what follows we provide a full derivation of the dispersion equation (7) that is used to calculate the photonic modes in the bulk metamaterial. This equation can be obtained by substituting the effective permittivity and permeability tensors of the metamaterial into the Maxwell equations. In that case the Maxwell equations read:

$$\nabla \times \mathbf{E} = i\omega \bar{\bar{\mu}}_{\text{eff,A+B}} \cdot \mathbf{H} \qquad (B1)$$

$$\nabla \times \mathbf{H} = -i\omega \bar{\bar{\varepsilon}}_{\text{eff,A+B}} \cdot \mathbf{E} \qquad (B2)$$

Combining Eqs. (B1) and (B2) yields:

$$\left( \nabla \times \bar{\bar{\mu}}_{\text{eff,A+B}}^{-1} \cdot \nabla \times \bar{\bar{I}} - \omega^2 \bar{\bar{\varepsilon}}_{\text{eff,A+B}} \right) \cdot \mathbf{E} = 0 . \qquad (B3)$$

We are interested in the plane-wave solutions ($\nabla = i\mathbf{k}$) of this equation for propagation in the *xoz*-plane, so that $k_y = 0$. They are obtained by setting the determinant of the matrix in (B3) to zero. It can be checked that for $\bar{\bar{\varepsilon}}_{\text{eff,A+B}}$ given by (2) and $\bar{\bar{\mu}}_{\text{eff,A+B}}$ given by (A3), the nontrivial solutions of (B3) separate into TE ($\mathbf{E} = E_y \hat{\mathbf{y}}$ and $H_y = 0$) and TM ( $\mathbf{H} = H_y \hat{\mathbf{y}}$ and $E_y = 0$ ) polarizations.

After straightforward manipulations it is found that for TM eigenwaves, the solutions of Eq. (B3) satisfy the following homogeneous system of equations

$$\begin{pmatrix} k_z^2 - \left(\dfrac{\omega}{c}\right)^2 \dfrac{\varepsilon_{xx}}{\varepsilon_0} & 0 & -k_x k_z \\ 0 & -\left(\dfrac{\omega}{c}\right)^2 \dfrac{\varepsilon_{yy}}{\varepsilon_0} & 0 \\ -k_x k_z & 0 & k_x^2 - \left(\dfrac{\omega}{c}\right)^2 \dfrac{\varepsilon_{zz}}{\varepsilon_0} \end{pmatrix} \cdot \mathbf{E} = \left[ k^2\left(\overline{\overline{\mathbf{I}}} - \hat{\mathbf{u}}_y \hat{\mathbf{u}}_y\right) - \mathbf{kk} - \dfrac{\overline{\overline{\varepsilon}}_{\text{eff,A+B}}}{\varepsilon_0}\left(\dfrac{\omega}{c}\right)^2 \right] \cdot \mathbf{E} = 0$$

, (B4)

where $k^2 = k_x^2 + k_z^2$ and $\mathbf{kk} = \begin{pmatrix} k_x^2 & 0 & k_x k_z \\ 0 & 0 & 0 \\ k_x k_z & 0 & k_z^2 \end{pmatrix}$ is calculated for $k_y = 0$. Equation (B4) may also be written as:

$$\left[ \left( k^2\left(\overline{\overline{\mathbf{I}}} - \hat{\mathbf{u}}_y \hat{\mathbf{u}}_y\right) - \mathbf{kk} \right) \cdot \left( \dfrac{\overline{\overline{\varepsilon}}_{\text{eff,A+B}}}{\varepsilon_0} \right)^{-1} - \overline{\overline{\mathbf{I}}}(\omega/c)^2 \right] \cdot \mathbf{E} = 0. \quad (B5)$$

The bulk eigenmodes supported by the metamaterial are then determined by the nontrivial solutions of the homogeneous system of equations (B5), which are formally equal to the solutions of Eq. (7), the characteristic equation.

## Appendix C: The quality factor of the Fano resonance

In the following, we describe the formalism used to determine the $Q$-factor of the Fano resonances in a metamaterial slab. The $Q$-factor of a resonator at a resonant frequency $\omega_0$ may be defined as $Q = \omega_0 U/P$ [69], where $U$ is the electromagnetic energy stored in the resonator and $P$ is the total power loss in the resonator (either by radiative or dissipative processes). The $Q$-factor of a resonator may also be calculated from its eigenmodes with frequency $\tilde{\omega} = \tilde{\omega}' + i\tilde{\omega}''$ using $Q = -\tilde{\omega}'/(2\tilde{\omega}'')$ [70]. Note that even when the materials are lossless the $Q$-factor can be finite due to energy leakage. Hence, the $Q$-factor has the physical meaning of a measure of the lifetime of the eigenmode in the resonator. We calculate the *Q*-factor of the Fano resonance in the metamaterial slab from the supported eigenmodes. The eigenmodes of the metamaterial can be obtained by expanding the electromagnetic field in the air and metamaterial regions in terms of plane waves as in Eqs. (8)-(10), but in the absence of the incident wave. By imposing the same boundary conditions at both interfaces as in the scattering problem (Eqs. 11a-b and 13a-b) we obtain a homogeneous 8×8 linear system of equations, whose kernel determines the characteristic equation that gives the dispersion of the natural modes of oscillation for TM waves. The solution of the characteristic equation corresponds to the eigenfrequency $\tilde{\omega} = \tilde{\omega}' + i\tilde{\omega}''$ of the natural mode of oscillation and allows determining the *Q*-factor of the Fano resonance using $Q = -\tilde{\omega}'/(2\tilde{\omega}'')$ [70].